\documentclass[a4paper]{article}
\usepackage{amsmath,amssymb,amscd,epsfig,tabularx,subfigure}

\def\beq{\begin{eqnarray}}
\def\eeq{\end{eqnarray}}
\def\bsp{\begin{split}}
\def\esp{\end{split}}

\newcommand{\mf}[1]{{\mathfrak #1}}

\newcommand{\mb}[1]{{\mathbb #1}}

\begin{document}
\title{Discrete Symmetries in Translation Invariant Cosmological Models}
\author{Sigbj\o rn Hervik\footnote{e-mail:S.Hervik@damtp.cam.ac.uk}\\
DAMTP, Cambridge University\\
Wilberforce Rd.\\
Cambridge CB3 0WA, UK}

\maketitle

\begin{abstract}
In this paper we investigate a class of $(d+1)$ dimensional
cosmological models with a cosmological constant possessing an
$\mb{R}^d$ simply transitive symmetry group and show that it can be
written in a form that manifests the effect of a permutation
symmetry. We investigate the solution
orbifold and calculate the probability of a certain number of
dimensions that will expand or contract. We use this to
calculate the probabilities up to dimension $d=5$.
\end{abstract}

\section{Introduction}
The effect of symmetries of differential equations on their
solutions were first truly recognized by the Norwegian mathematician
Sophus Lie more than hundred years ago. Lie made considerable progress on
the effect of so-called
transformation groups on the solutions of differential equations. A
simple symmetry principle can yield many interesting properties of the
solutions to the equations of motion for a physical system. In gauge
theories the complete Lagrangian can be deduced by a requirement that
it should be invariant under a certain
symmetry\footnote{One usually also demands that the terms in the
Lagrangian should be renormalisable.}. In cosmology there have been
many studies written of the so-called Bianchi
universes\footnote{See for instance \cite{RS}.}. Bianchi
universes are spatially homogeneous cosmological models that can be
classified according to Bianchi's classification of the 3-dimensional
Lie algebras. They are numbered $I-IX$ and are in general anisotropic.

The object of this paper is to investigate a certain family of
spatially homogeneous cosmological models. We will investigate $(d+1)$ dimensional space-times with a simply transitive symmetry group $\mb{R}^d$. The metric for these models can be written 
\beq\label{bt1d}
ds^2=-N(t)^2dt^2+\sum_{i=1}^d a_i(t)^2(dx^i)^2
\eeq
For $d=3$ this is called the Bianchi type I model. In this paper we
will focus on another symmetry that these space-times possess. We
note that the metric \ref{bt1d} is also
invariant under the discrete symmetry group $S_d$, {\it the symmetric
group}, or {\it the permutation group of $d$ elements}. The labelling of each coordinate $x^i$ , $i=1,...,d$ is somewhat
artificial and can be permuted to any other sequence, thus the mapping
$(x^1, ... ,x^i,... x^d)\longmapsto (x^{k_1}, ... ,x^{k_i},... x^{k_d})$
where $(k_1, ... ,k_i,...k_d)$ is any permutation of $(1,...,i ,...,
d)$ is a symmetry transformation. This will be the main observation
of this paper. 

Let us introduce the notion of a regular $(d-1)$-simplex,
$\sigma_{d-1}$. It is the generalization of an equilateral triangle to
any dimension. We embed $\sigma_{d-1}$ in Euclidean $(d-1)$ space,
$\sigma_{d-1} \subset \mb{R}^{d-1}$ and write ${\bf n}_i$ to denote
the position of the $i$th vertex relative to the center of mass
frame. For simplicity, if we write ${\bf n}_i$ as a column vector, we
define a $(d-1)\times d$ matrix ${\bf Q}$ by 
\beq
{\bf Q} =[{\bf n}_1,...,{\bf n}_d]= [ (n^a_i)]
\eeq
By regularity of $\sigma_{d-1}$, we get the following two relations
\beq
\sum_{i=1}^d {\bf n}_i &=&0 \\
|{\bf n}_i-{\bf n}_j|&=&\ell ,~ i\neq j
\eeq
We can now state the theorem which we shall prove in the next section:
\paragraph{Theorem:}
{\it The general solution for the line element (\ref{bt1d}) of the
vacuum Einstein field equations $R_{\mu \nu}-\frac{1}{2}Rg_{\mu \nu}+\Lambda g_{\mu \nu}=0$ in $(d+1)$ dimensions is 
\beq\label{solution}
ds^2=-\frac{dt^2}{\left( 1-\frac{d}{2(d-1)}\Lambda t^2\right)^2}+\frac{t^{\frac{2}{d}}}{\left( 1-\frac{d}{2(d-1)}\Lambda t^2\right)^{\frac{2}{d}}}d\tilde{s}^2
\eeq
where 
\beq\label{solution2}
d\tilde{s}^2=\sum_{i=1}^d t^{\frac{2(d-1)}{d}n^a_i}(dx^i)^2
\eeq
and $n^a_i$, $i=1,...,d$ correspond to a regular $(d-1)$-simplex
inscribed in the unit $(d-2)$ sphere, or de Sitter's solution
with flat spatial sections. }
\par
\begin{figure}
\centering
{\epsfig{figure=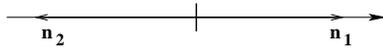, width=5.5cm} 
\caption{The $d=2$ case. } 
\label{d2}}
\end{figure}
\begin{figure}
\centering
{\epsfig{figure=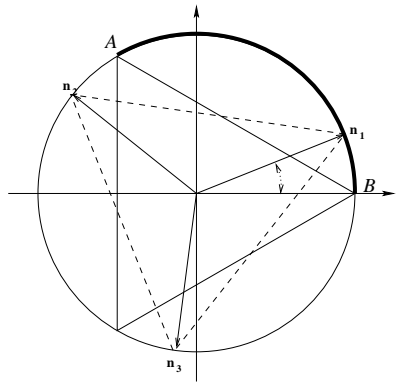, width=5.5cm} 
\caption{The $d=3$ case. The bold arc $AB$ is a double covering
of the solution orbifold in this case. The point represented by the
angle $\frac{\pi}{3}$ will be a reflection point under the
orbifold identification.} 
\label{d3}}
\end{figure}
\begin{figure}
\centering
{\epsfig{figure=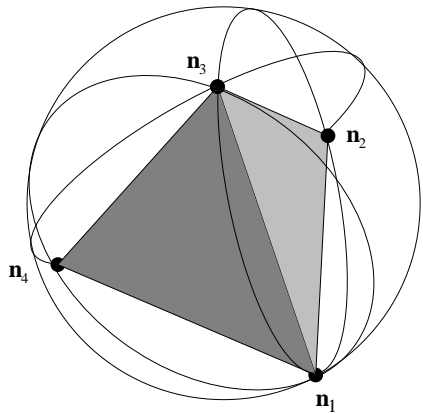, width=5.5cm} 
\caption{The $d=4$ case. } 
\label{d4}}
\end{figure}

De Sitter's solution is, as we shall see, completely disconnected from the
other solutions and we will in the further disregard this solution. 
This theorem has been proven for $d=2$ and $d=3$ in previous
works \cite{Carlip,MyPaper}. The $d=2$ case is illustrated in figure \ref{d2}. The $d=3$
case which is usually called the Bianchi type I solution, is proved in
\cite{MyPaper} and illustrated in figure \ref{d3}. We have also
illustrated the case $d=4$ in figure \ref{d4} which is
a regular tetrahedron inscribed in the unit
2-sphere. 

The case $d=1$ is trivial and a 0-simplex is simply a point. But since the ${(-1)}$-sphere is somewhat poorly defined, we can just say that $\sigma_0={0}$. This results in the Milne line element for $\Lambda=0$:
\[ ds^2=-dt^2+t^2dx^2 \]
We will exclude the $d=1$ case in what follows.

\section{Proof}
For simplicity we will consider a vanishing cosmological constant. The results can easily be expanded to a non-zero cosmological constant by similar methods as in the paper \cite{MyPaper}. 

With $\Lambda=0$ one can easily derive the solutions of
eq. (\ref{bt1d}). The solutions are usually written\footnote{For
simplicity we assume that indices like $i,j,..$ have range $1,...,d$ while indices like $a,b,... $ have range $1,...,d-1$.}:
\beq 
ds^2=-dt^2+\sum_{i} t^{2p_i}(dx^i)^2
\eeq
where $\sum_{i}p_i=1$ and $\sum_{i}p_i^2=1$.

By using of the permutation group $S_d$ under which the solution space must be invariant, we will show that these two descriptions are equivalent. 
We first inscribe the regular simplex $\sigma_{d-1}$ in the unit sphere $S^{d-2}$. Since ${\bf n}_i\in S^{d-2}$, we have $\sum_{a}(n_i^a)^2=1$. By evaluating the sums $\sum_{i,j}\sum_a (n^a_i-n_j^a)^2$ and $\sum_a (n_i^a -n_j^a)^2$ in two different ways we can obtain the following useful relations:
\beq
\ell &=& \sqrt{\frac{2d}{d-1}} \\
{\bf n}_i\cdot {\bf n}_j|_{i\neq j} &=& -\frac{1}{d-1} \label{dotprod}
\eeq

We notice that the symmetry group of a regular $(d-1)$-simplex Sym($\sigma_{d-1}$) (including orientation-reversing operations) is isomorphic to $S_d$. More specifically, each element of the group Sym($\sigma_{d-1}$) is a permutation of the vertices ${\bf n}_i$. Thus we immediately have two different representations of the permutation group $S_d$. The first is $d\times d$ matrices $\Gamma(g)\in O(d)$ ($g\in S_d$)\footnote{$O(n)$ the is matrix group of orthogonal matrices: $O(n)=\{A|A^TA=AA^T=1\}$ }. These matrices represent the permutation of the vectors ${\bf n}_i$. The second representation consists of $(d-1)\times (d-1)$ matrices $\gamma(g)\in O(d-1)$ and the matrices $\gamma(g)$ are for a fixed orientation of $\sigma_{d-1}\subset \mb{R}^{d-1}$ the $ O(d-1)$-rotations of $\sigma_{d-1}$ that map $\sigma_{d-1}$ onto itself. Hence we have the following relation:
\beq 
\gamma(g){\bf Q}={\bf Q}\Gamma(g)
\eeq
Notice that along the diagonal of the matrix ${\bf N}\equiv {\bf Q}{\bf Q}^T$ we have elements of the type $\sum_i (n_i^a)^2$. We further note that 
\beq
\gamma(g){\bf N}\gamma(g)^T=\gamma(g){\bf Q}{\bf Q}^T\gamma(g)^T={\bf
Q}\Gamma(g)\Gamma(g)^T{\bf Q}^T={\bf Q}{\bf Q}^T={\bf N}
\eeq
for all $g\in S_d$. On the left side of this equation the diagonal elements are permuted compared to the right side. Since the above equation holds for all $g\in S_d$ we must have 
\beq 
\sum_i (n_i^a)^2=\sum_i (n_i^b)^2, ~\forall a,b
\eeq
We now obtain
\beq
\sum_i(n^a_i)^2=\frac{1}{d-1}\sum_a\sum_i(n^a_i)^2=\frac{1}{d-1}\sum_i\sum_a(n^a_i)^2=\frac{1}{d-1}\sum_i 1 =\frac{d}{d-1}
\eeq
Now we show that $\sum_i(\frac{1}{d}+\frac{d-1}{d}n^a_i)$ and $\sum_i(\frac{1}{d}+\frac{d-1}{d}n^a_i)^2$ both sum to 1. The first sum is
\beq 
\sum_i\left(\frac{1}{d}+\frac{d-1}{d}n^a_i\right)&=&\frac{1}{d}\sum_i1+\frac{d-1}{d}\sum_in^a_i\\ &=&1;
\eeq
while the second sum is
\beq
\sum_i\left(\frac{1}{d}+\frac{d-1}{d}n^a_i\right)^2&=&\frac{1}{d^2}\sum_i\left[1+2(d-1)n^a_i+(d-1)^2(n^a_i)^2\right]\\ &=& \frac{1}{d^2}[d+(d-1)d]\\ &=&1
\eeq
Thus we have shown that equations (\ref{solution}) and (\ref{solution2}) are solutions of the Einstein field equation as the theorem claims. But does the theorem span the whole set of solutions that can be written $\sum_{i}p_i=1$ and $\sum_{i}p_i^2=1$? The equation $\sum_{i}p_i^2=1$ describes a $(d-1)$ sphere and $\sum_{i}p_i=1$
a $(d-1)$ dimensional hyperplane. The intersection of these two spaces
is a $(d-2)$ sphere $S^{d-2}$. From the other
perspective, the group $G=O(d-1)$ acts transitively on the sphere
$S^{d-2}$ and thus is the group of all orientations of the regular
simplex $\sigma_{d-1}$. The solution space of the metrics given in the
theorem is the projection of the vertices of $\sigma_{d-1}$. This
projection onto (for instance) the 1st axis is invariant under the
isotropy group of $(1,0,...,0)\in S^{d-2}$ which is $O(d-2)$. Thus the
solution space of the metric of the theorem has the universal
cover\footnote{This relation can for instance be found in
\cite{Nakahara}} $O(d-1)/O(d-2)\cong S^{d-2}$ . Thus the two
descriptions have  solution spaces which equal the same
compact space. Hence they are equivalent.

We would also emphasize that the solution space is the orbifold
$S^{d-2}/S_d$, i.e. we identify points on the $(d-2)$-sphere under the
action of the symmetric group. This action is not free: there
are fixed points under this action. If $y\in S^{d-2}$ is fixed under
a subgroup $G\subset S_d$ then we say that $y$ is a $G$ fixed
point. $G$ will be isomorphic to $S_h$ for\footnote{Equality is
only allowed for isotropic FRW models.}  $h\leq d$ or a direct product
of these, and if $y$ is a $S_{h_1}\times ... \times S_{h_m}$ fixed
point then $y$ will correspond to a $SO(h_1)\times
... \times SO(h_m)$ symmetric spacetime. Hence these fixed points are
actually spacetimes with higher symmetry than originally assumed. 

\subsection*{The inversion map}
It is quite clear that the mapping: ${\mathcal R}: {\bf Q} \longmapsto -{\bf Q}$ maps a solution ${\bf Q}$ onto an other solution $-{\bf Q}$. Geometrically this is the reflection of the simplex through the origin. This is not one of the symmetry operations of the regular simplex, thus $-{\bf Q} \neq {\bf Q}$. 

Let us simultaneously consider the mapping:
\beq
\mf{I}: t\longmapsto \frac{1}{\omega^2 t}
\eeq
for $\Lambda\neq 0$ and $\omega^2=\frac{d}{2(d-1)}|\Lambda|$. The map
$\mf{I}$ maps expanding solutions onto contracting solutions. For
positive $\Lambda$ this means that the contracting branch, where
$\omega t>1$, is mapped onto the expanding branch $\omega t<1$. We
note that the composition of the maps ${\mathcal R}$ and $\mf{I}$ is
almost the identity map (or an isometry). The resulting metric can be
written in the same way as the original metric except that the spatial coordinates have been rescaled. The resulting map 
\beq
dx^i \longmapsto \omega^{\frac{2(d-1)}{d}n^a_i} dx^i
\eeq
is just a rescaling of the space coordinates. It also interesting to
note when this mapping is an isometry. This always happens if $\omega=1$. If $\omega\neq 1$ then either $n^a_i=0$ or the spatial coordinates $x^i$ have to be infinite in range. This has to be true for all $i$. 

Even though the map $\mf{I}$ is not defined for $\omega=0$ the map $\mathcal R$ is, and we can say that the map $\mathcal R$ is a duality relation between contracting and expanding solutions. 

\section{A special class of solutions: flat-fibered FRW solutions}

In the previous section we have given a simple relation fulfilled by
any $d$-dimensional, translation invariant solution. In this
section we will look at a subclass of solutions that
describe a flat FRW universe with flat fibres, and we will again
assume that $\Lambda\neq 0$. 

\begin{figure}
\centering
{\epsfig{figure=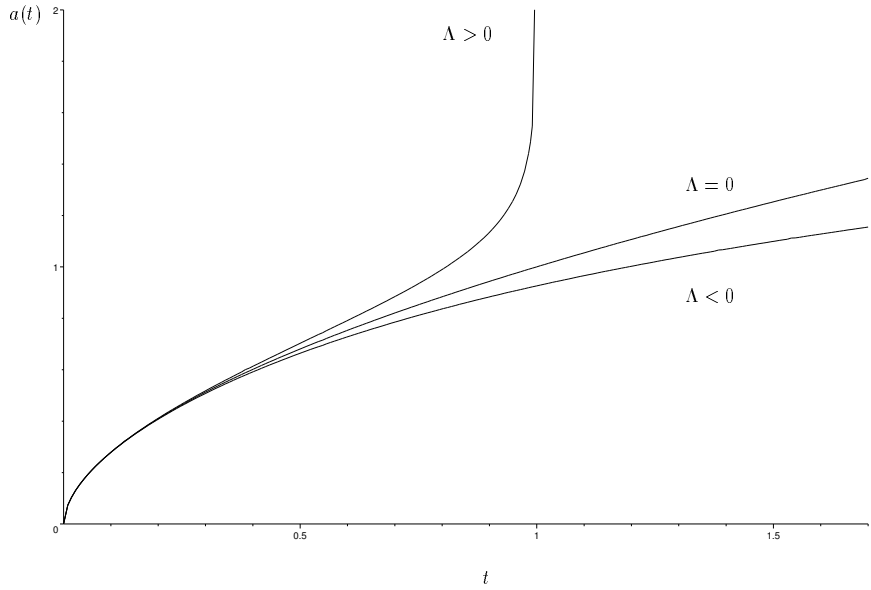, width=12cm} 
\caption{The typical expansion of a scale factor in the expanding part
of the metric in the Kasner time gauge. For $\Lambda >0$ the scale
factors will go to infinity as $t\longrightarrow \omega^{-1}$ because
of the denominator in (\ref{scalefactors}). The time $t=\omega^{-1}$
represents the isotropic limit. } 
\label{Kasnergauge}}
\end{figure}

\begin{figure}
\centering
{\epsfig{figure=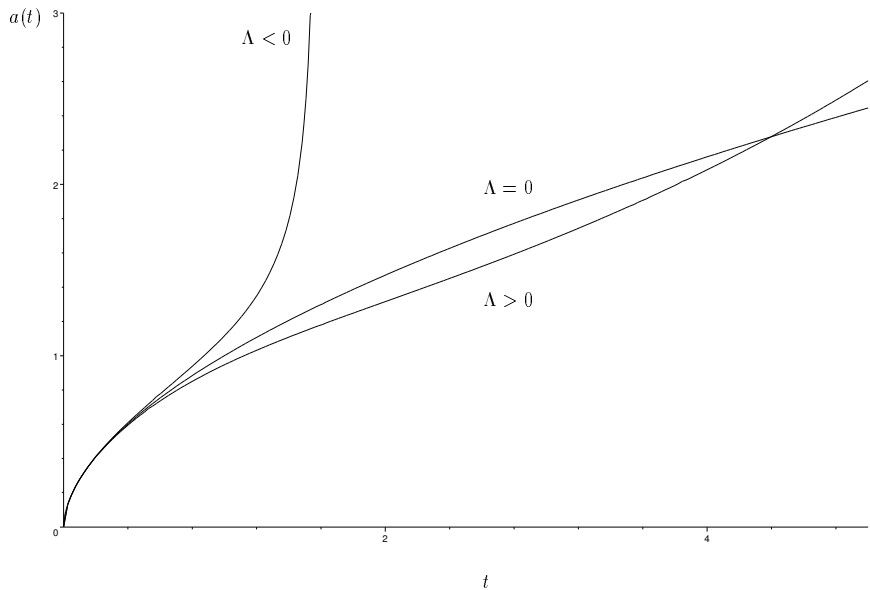, width=12cm} 
\caption{The typical expansion of a scale factor in the expanding part
of the metric in the universal time gauge. Note that as the $d$-volume
collapses in the final singularity, the expanding part will grow to
infinite size for $\Lambda<0$.} 
\label{universalgauge}}
\end{figure}

In the previous sections we have used a time gauge in which each of
the scale factors $a_i(t)\equiv \sqrt{g_{ii}}$ are (up to a
proportionality factor) given by
\beq\label{scalefactors}
a_i(t)=\frac{t^{\frac{1}{d}(1+(d-1)n_i^a)}}{(1-\lambda t^2)^{\frac{1}{d}}}
\eeq
This time gauge will be called the Kasner time
gauge. Notice that the spatially $d$-volume varies as 
\beq
V(t)\propto \prod_{i=1}^d a_i(t)=\frac{t}{1-\lambda t^2}
\eeq
 We define the universal time gauge as 
\beq
\eta=\int_0^t \frac{d\tau}{1-\lambda \tau^2}
\eeq
The scale factors in the universal time gauge are ($\omega=\sqrt{|\lambda|}$)
\beq
a_i(\eta)=\begin{cases}
(\sinh\omega \eta)^{\frac{1}{d}(1+(d-1)n_i^a)}(\cosh\omega\eta)^{\frac{1}{d}(1-(d-1)n_i^a)}, & \Lambda> 0 \\
\eta^{\frac{1}{d}(1+(d-1)n_i^a)}, & \Lambda=0 \\
(\sin\omega \eta)^{\frac{1}{d}(1+(d-1)n_i^a)}(\cos\omega\eta)^{\frac{1}{d}(1-(d-1)n_i^a)}, & \Lambda< 0 
\end{cases}
\eeq
while the spatially $d$-volume is
\beq
V(\eta)\propto \prod_{i=1}^d a_i(\eta)=\begin{cases}
(\sinh\omega \eta)(\cosh\omega\eta), & \Lambda> 0 \\
\eta, & \Lambda=0 \\
(\sin\omega \eta)(\cos\omega\eta), & \Lambda< 0 
\end{cases}
\eeq
Let us define the following two vectors for positive integers $n,m$ where $n+m=d$:
\beq
\xi_n=\frac{1}{n}\sum_{i=1}^n {\bf n}_i
\eeq
and
\beq
\xi_m=\frac{1}{m}\sum_{i=1+n}^{m+n} {\bf n}_i
\eeq
Their norms are easily calculated by the use of $|{\bf n}_i|=1$ and relation \ref{dotprod}:
\beq
|\xi_n|^2=\frac{m}{n(d-1)} \\
|\xi_m|^2=\frac{n}{m(d-1)}
\eeq
Assuming that $n_i^a=|\xi_n|$ for $1\leq i \leq n$ and
$n_i^a=-|\xi_m|$ for\footnote{Note that this
implies that we are considering a $S_n\times S_m$ fixed point that
will correspond to a $SO(n)\times SO(m)$ symmetric spacetime.}
$1+n\leq i \leq n+m=d$, it follows that  $\sum_i(n^a_i)^2 =\frac{d}{d-1}$ and $\sum_i n^a_i =0$. A interesting point is that the $n^a_i$'s are bounded by
\beq
(d-1)^2\geq (d-1)|n^a_i|\geq 1
\eeq
Thus we will have
\beq
\dot{a}_i
\geq 0 & , \text{for} ~ 1\leq i\leq n 
\eeq
independent of the value of the cosmological constant (and whether
$a_i$ is expressed in the universal time gauge or in the Kasner
gauge). Thus we have that one flat section is always expanding, even if the cosmological
constant is negative!(see  figures \ref{Kasnergauge} and
\ref{universalgauge}.)

\subsection*{An example from String Theory}
All the five consistent string theories demand a (9+1) dimensional
spacetime\footnote{See for instance \cite{Polchinski}.}. One usually
compactifies 6 spatial dimensions to a Calabi-Yau (CY)
manifold\cite{Green}. If $d\Sigma_m^2$ is the $m$-dimensional
Euclidean metric we will assume therefore that the metric takes the form
\[ ds^2=-N^2dt^2+a^2d\Sigma_3^2+a_{CY}^2d\Sigma_6^2. \]
This metric describes a flat FRW metric with a 6-dimensional fibering. We now take a 8-simplex and inscribe it in the unit 8-sphere, assuming that $n_1^1=n_2^1=n_3^1$ and $n_4^1=...=n_9^1$. We then find:
\beq 
|\xi_3|=\frac{1}{2}
\eeq
Similarly, we have
\beq 
|\xi_6|=\frac{1}{4}.
\eeq
If we set $n_i^1=|\xi_3|$ and $n_j^1=-|\xi_{6}|$ for $i=1,2,3$ and $j=4,...,9$ the $n_k^1$ fulfill all the necessary relations. 
Thus we get the solution:
\beq
ds^2=-\frac{dt^2}{\left( 1-\frac{9}{16}\Lambda t^2\right)^2}+\frac{1}{\left( 1-\frac{9}{16}\Lambda t^2\right)^{\frac{2}{9}}}\left(t^{\frac{10}{9}}d\Sigma_3^2+t^{-\frac{2}{9}}d\Sigma_6^2\right)
\eeq
Let us now assume $\Lambda=0$. 
If we compactify $d\Sigma_6^2$ to be a compact CY-manifold\footnote{The simplest of these is the 6-dimensional torus $T^6$.} and compactify\footnote{Here there are 6 different orientable possibilities, again the simplest is the torus $T^3$. The other 5 can be written as a quotient of $T^3$.} $d\Sigma_3^2$ we note that while the volume of $d\Sigma_3^2$ is expanding, Vol$(d\Sigma_3^2)\propto t^{\frac{5}{3}}$, the volume of the CY is contracting, Vol$(d\Sigma_6^2)\propto t^{-\frac{2}{3}}$. Thus even in this simple classical model we obtain a mechanism that yields a possibility that while the physical 3-space expands, the CY-manifold shrinks to arbitrarily small size. If we had included a cosmological constant the relative sizes between the 3-space and the CY would become constant after a while. Thus if this universe undergoes inflation, the CY manifold would be fixed to a small size compared to the other 3 spatial dimensions.

The other solution where the spacetime metric possesses the same symmetries as the above metric is found by using the reflection map $\mathcal R$. The result is simply:
\beq
ds^2=-\frac{dt^2}{\left( 1-\frac{9}{16}\Lambda t^2\right)^2}+\frac{1}{\left( 1-\frac{9}{16}\Lambda t^2\right)^{\frac{2}{9}}}\left(t^{-\frac{2}{3}}d\Sigma_3^2+t^{\frac{2}{3}}d\Sigma_6^2\right)
\eeq
Here we have an opposite behaviour, the volume of $d\Sigma_3^2$ is
contracting as Vol$(d\Sigma_3^2)\propto t^{-1}$, and the volume of the
CY is expanding Vol$(d\Sigma_6^2)\propto t^{2}$. We note a
peculiar thing, the effect on the FRW part of the ``reflected'' solution
of the model has a different behaviour than the original one. This is
directly related to the fact that the dimensionalities of the FRW part
and CY part are not equal. 

\section{Universes with matter and investigations of the solution orbifolds}
So far we have discussed vacuum solutions only with a
cosmological constant. We have investigated some special classes of
solutions but have not actually discussed whether these solutions are
probable or not. In the following we will introduce matter into
the universe, a special kind of matter that possesses similar
symmetries as that of the space itself. Let us look at scalar fields
with a Lagrangian:

\beq
{\mathcal L}_M=-\frac{1}{2}\sqrt{-g}\left[ \sum_{c=1}^{\delta} \partial_{\mu}\phi_c
\partial^{\mu}\phi_c+ V(\phi_c)\right]
\eeq
where $\delta$ is the number of scalar fields. 
To sustain spatial homogeneity in our models the scalar fields have to
be position independent i.e.
$\phi_c(t)$. Thus if $V(\phi_c)$ is constant\footnote{This constant is
equivalent to introducing a cosmological constant and so we will
assume that this constant is zero.} the scalar fields will
possess a $\mb{R}^{\delta}$ symmetry. These scalar
fields can play the role of additional dimensions in these
models. With the aid of the Lagrangian this is quite easy to see. The
gravitational Einstein-Hilbert action $S_G=\int_M\sqrt{-g}(R-2\Lambda)d^4x
+\int_{\partial M} 2K d^3x $ turns out to be:
\beq 
S_G=\int
dt\left[\frac{e^{d\alpha}}{N}\left(-d(d-1)\dot{\alpha}^2+\sum_{i=1}^d
\dot{\beta}_i^2\right)-2N\Lambda e^{d\alpha}\right]
\eeq 
where $a_i\equiv \exp(\alpha+\beta_i)$ and $
\sum_{i=1}^d{\beta_i}=0$. Since the $\beta_i$'s are not
independent we will instead introduce a set of variables $X^a,~
a=1,...,d-1$. We choose a simplex as described in earlier sections
with the corresponding matrix ${\bf Q}$, we can write both
$(\beta_i)$ and $(X^a)$ as column vectors, with
\beq 
\beta=(d-1){\bf Q}^T X
\eeq
This is a linear homeomorphism and the image is
exactly the set of allowed $\beta_i$'s. This can be seen if one notes
that the matrix ${\bf N}={\bf Q}{\bf Q}^T$ is proportional to the identity matrix
(due to Schur's lemma). Then, we have 
\beq
\sum_{i=1}^d\dot{\beta}_i^2=d(d-1)\sum_{a=1}^{d-1}(\dot{X}^a)^2
\eeq

Thus the action for a $d$-dimensional universe with $\delta$ massless
scalar fields is simply
\beq
S=\int
dt\left[\frac{e^{d\alpha}}{2N}\left(-2d(d-1)\dot{\alpha}^2+
2d(d-1)\sum_{i=1}^{d-1}(\dot{X}^a)^2+\sum_{c=1}^{\delta} \dot{\phi}_c^2
\right) -2N\Lambda e^{d\alpha}\right]
\eeq 
 Hence up to a trivial rescaling of the $\phi_c$, the $X^a$ and the
$\phi_c$ appear in the action on equal footing. The action cannot
see the difference between the scalar field and the $X^a$
\emph{except} in the global scaling of the volume factor $\alpha$. 

The classical solutions can now be obtained. Choosing the
Kasner time gauge $\exp(d\alpha )= t N$ the solutions are (up to a constant by addition\footnote{In the case where the
universe is compactified these constants cannot be set to zero by an
isometry. The constants will then represent scaling parameters in the
different compact directions. }):
\beq
X^a&=&\frac{1}{d} \eta^a \ln t \\
\phi_c&=&\sqrt{\frac{2(d-1)}{d}}\eta^{d-1+c}\ln t \\
\exp(d\alpha)&=& \frac{t}{1-\frac{d}{2(d-1)}\Lambda t^2}
\eeq
Writing $\eta^{\mu}, ~\mu = 1,..., d+\delta-1$ the $\eta^{\mu}$ will be
coordinates on the unit $(d+\delta -2)$-sphere: $\sum_{\mu}(\eta^{\mu})^2=1$.
The scalar fields behave just as the other variables $X^a$ and is in
some sense compensating for the other dimensions that are not there!
The scalar fields are reducing
the sphere containing the simplex. If the sphere has zero
radius the metric is a FRW metric with no anisotropy. All the $\eta^{\mu}$'s
for $\mu=1,...,d-1$ have to be exactly zero and the result turns
simply into the case where 

\beq
d\tilde{s}^2= \sum_{i=1}^{d}(dx^i)^2
\eeq
However, this is an exceptional case. In the set of all configurations this
set of FRW universes\footnote{Actually the set of FRW universes
consists of only two elements.} is of zero measure, but the set of configurations 
where all Kasner parameters
are positive has non-zero measure. 

As we have seen, the vacuum Bianchi type I solutions $(d=3)$ can be
seen as the orbifold $S^1/S_3$. Since the order of $S_3$ is 6, which
is usually written $|S_3|=6$, the solution orbifold can be seen as an
arc of length $\frac{\pi}{3}$. The fixed points under the orbifold
identification are the end points of this arc and
represent universes with higher symmetry than originally assumed. In
the $d=3$ case these points correspond to the two different $SO(2)$ symmetric
spacetimes. An  $SO(2)$ symmetric
spacetime has zero measure. All the other vacuum
solution cases will have 2 expanding ($e$) and 1 contracting ($c$) direction,
i.e. $P(2e+1c)=1$ where $P$ is the probability. If we include a scalar
field the situation changes to
$S^2/S_3$. Now we assume that the probability
of each solution is weighted by the natural geometry of
the solution orbifold (which is spherical). The probabilities turn out to be
\beq 
P(2e+1c)&=& \frac{3}{4} \\
P(3e)& =& \frac{1}{4}
\eeq
The most probable scenario is still a universe with 2 expanding and 1
contracting solution but nevertheless there is a non-zero probability
that there are 3 expanding directions.

\subsection*{The case $d=4$} 
Increasing the number of dimensions the complexity increases
drastically. However the next step, the $d=4$ case, can 
also be illustrated easily. 

\begin{figure}
\centering
{\epsfig{figure=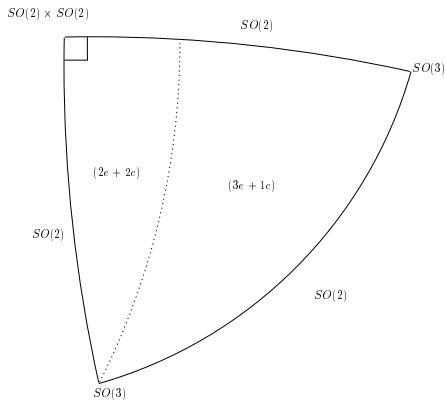, width=6cm} 
\caption{The vacuum $d=4$ solution orbifold. The dotted line separates the
$(2e+2c)$ solutions from the $(3e+1c)$ solutions.} 
\label{d4orbi}}
\end{figure}

The solution space of the Kasner coordinates can now be
illustrated by a $(\frac{\pi}{2}, \frac{\pi}{3}, \frac{\pi}{3})$
spherical triangle on the 2-sphere $S^2$, see figure \ref{d4orbi}. 
Again, all the fixed points
are on the boundary of this triangle and correspond to universes with
higher symmetry than originally assumed. The $\frac{\pi}{2}$ vertex
correspond to a universe with symmetry $SO(2)\times SO(2)$ and the
other 2 vertices correspond to the two different $SO(3)$ symmetric
spacetimes. The rest of the rim of the triangle represents different versions
of $SO(2)$ symmetric spacetimes. But these spacetimes
 have measure zero as a set compared to the whole solution
set. 

In this case it is also interesting to calculate the various
probabilities of different numbers of contracting and expanding directions. The result for the
vacuum case is:
\beq 
P(2e+2c)&=& \frac{1}{3} \\
P(3e+1c)&=& \frac{2}{3}
\eeq
Now we see that the odds for a universe with 3 expanding directions
are 2:1. Thus the most probable case is a universe with 3 expanding
directions and 1 contracting one. 

The case with 3 expanding directions will be even more probable if
we include a scalar field. Then the solution space
is that of a 3-sphere $S^3$ with certain identifications under the
symmetric group $S_4$. Interestingly, using the orthogonal projection,
we can map this space onto two solid balls, each ball corresponds to the
North and South hemispheres of $S^3$. By suitable choice of coordinates,
one ball is for positive scalar field, the other for negative. The
characteristic solutions will now divide the space into different
regions. Except for sets of measure zero, these regions will correspond
to $(2e+2c)$, $(3e+1c)$ and $(4e)$ solutions. In the
orthogonal projection model, the $(4e)$ regions will be two solid regular
ideal tetrahedrons, one in each ball. However, using the pull-back
metric from the projection of the sphere $S^3$, the metric in the two balls
will be:
\beq
ds^2=\frac{dr^2}{1-r^2}+r^2(d\phi^2+\sin ^2 \phi d\theta^2)
\eeq
Finding the different probabilities will now be considerably more
difficult but can be calculated numerically. The result
is: 
\beq
P(2e+2c)&=& 0.136\\
P(3e+1c)&=& 0.805\\
P(4e) &=& 0.059
\eeq
 Thus, there will be with an 80\% chance
that the universe (at least initially) has 3 expanding directions and
1 contracting. One dimension will automatically contract and be of
arbitrary small size compared to the other ones. It is assumed that
quantum gravity effects dominates when $t<t_{Pl}$ and one would
naively believe that the wavefunction is evenly distributed over
configuration space. In the region $t>>t_{Pl}$ collisionless particles
counteract on the metric and make contracting directions
unstable\cite{LS}. The contracting directions will eventually start to expand
and the net effect is an evolution towards an isotropic universe.   

The $d=5$ vacuum case is also calculated and some explanation of the
calculation is made in the appendix. The results obtained
are summarized in table \ref{summary}.
\begin{table}
\centering
\setlength{\extrarowheight}{4pt}
\begin{tabular}{| >{\large}c || >{\large}c | >{\large}c| >{\large}c |
>{\large}c| >{\large}c |}
\hline
& $d=3$ V & $d=3$ SF & $d=4$ V&  $d=4$ SF& $d=5$ V \\
\hline \hline 
2e & 1 & $\frac{3}{4}$ & $\frac{1}{3}$ & 0.136 & 0.257 \\ \hline
3e & 0 & $\frac{1}{4}$ & $\frac{2}{3}$ & 0.805 & 0.389 \\ \hline
4e & - & - & 0                         & 0.059 & 0.355  \\ \hline
5e & - & - & - & - & 0 \\ \hline

\end{tabular}\\ 
\caption{Summary of the probability calculations done in this
paper. The V stands for vacuum case, the SF for scalar field case.}
\label{summary}
\end{table}

\section{Conclusion and Discussion} 
We have shown that the solutions of the general
$(d+1)$ dimensional spacetime with a translational invariant metric can be written in different ways. The
solution space of the model had the orbifold structure $S^{d-2}/S_d$
in the vacuum case and $S^{d-1}/S_d$ when a scalar field is present. We
wrote the solutions as a $(d-1)$-plet under the
permutation group which could be interpreted as the coordinates of the
vertices of a regular $(d-1)$-simplex. This simplex possesses the same
discrete symmetries as the model  and gave a geometrical description
of the solution space. 

This representation of the solution space was useful for several
reasons. As we saw in \cite{MyPaper}, we may have a continuous
transition to an isotropic FRW universe by a reduction of the radius
of the sphere in which the simplex $\sigma_{d-1}$ is inscribed. The
symmetries of the solutions are manifest under this continuous
transition. Another property of the ordinary Bianchi type I solutions
is that there are particular solutions which appear to be
special. These solutions play a very special role in the solution
space: They are the fixed points under the orbifold identification of
the sphere with respect to the symmetry group of the simplex which was isomorphic
to the symmetric group $S_d$. These fixed points have the special
property that they represent spacetimes that have a larger symmetry
group than the original $\mb{R}^d$. Also, the type of the fixed point,
(i.e. if it is a reflection fixed point or an $S_3$ fixed point)
determined the symmetry of the resulting spacetime. In general,
an $S_n$ fixed point would yield a $SO(n)$ symmetric spacetime and any
product $S_n\times ... \times S_m$ fixed point would yield a $SO(n)\times
... \times SO(m)$ symmetric spacetime (in addition to translation
symmetry). 
The set of solutions which is represented by ${\bf Q}$ does not have these
special points. This is because the solution space of different
${\bf Q}$'s are $O(d-1)$ and the representation of the group $S_d$ acts
freely and  properly discontinuous on the manifold $O(d-1)$ by left
(or right) multiplication. Thus the quotient $O(d-1)/S_d$ is a smooth
manifold \cite{thur97}. There are no special points. 

We also obtained a cosmological solution which describes a
(9+1) dimensional spacetime where 6 flat dimensions contract to
arbitrarily small size, while 3 spatial dimensions expand to arbitrarily
large size. However we emphasized that this is a special solution and
the set of such solutions had only measure zero in the whole solution.
By using this geometrical description we calculated various
probabilities that certain configurations existed. 

In this we paper have given a systematic approach to how such
solution spaces ``look like''. It would be interesting to finding out
more of the properties in even higher dimensions than those
investigated here. There may be a link between these solutions and the
concept of chaos which is known to exist in certain of these
spacetimes \cite{DH2000}. 

In the context of quantum cosmology these issues are indeed very
interesting. As we have in this paper only considered classical
probabilities under the assumption that all the points in the solution
space were equally probable, it would be interesting to investigate
whether the quantum calculations will yield the same results or
whether more symmetric cases are favoured \cite{CGS, GC}.

\section*{Acknowledgments}
I would like to thank both John D. Barrow and
\O yvind Gr\o n for their useful comments.
Part of this work was funded by the Research Council of Norway.

\section*{Appendix}

A $D$-simplex $\sigma_D\equiv [{\bf n}_1,...,{\bf n}_{D+1}]$ (the
order is irrelevant) can be parametrized in
$\mb{R}^m,~m\leq D$ by:
\beq
\sigma_D=\left\{{\bf x}\in \mb{R}^m|{\bf x}=\sum_{i=1}^{D+1}c_i{\bf n}_i,
c_i\geq 0, \sum_{i=1}^{D+1}c_i=1\right\} 
\eeq
A thing to note here is that a simplex is actually a whole set of
simplices:
\beq
K=\left\{ [{\bf n}_1], ... , [{\bf n}_{D+1}],[{\bf n}_1,{\bf n}_2],
... , \sigma_{D+1} \right\}
\eeq
In $K$ the number of $i$-simplices are $\frac{(D+1)!}{(D-i)!(i+1)!}$.

\begin{figure}
\centering
{\epsfig{figure=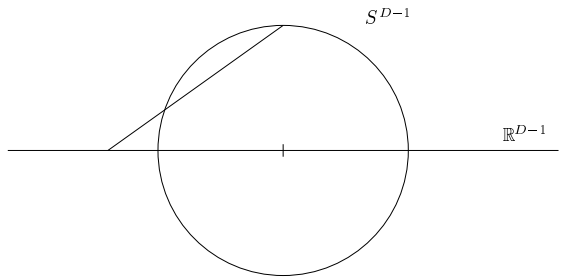, width=8cm} 
\caption{The stereographic projection of the sphere.} 
\label{stereo}}
\end{figure}

Let us consider a regular simplex with all its vertices on the unit
sphere: $\sigma_{D}\subset \mb{R}^D, ~ {\bf n}_i\in S^{D-1}$. The
(covering) solution space of the vacuum case is now represented by the
sphere $S^{D-1}$. Under the orthogonal projection the solution space
of the models with one scalar field will be two such spheres
$S^{D-1}$ but including their interior and with their boundaries
identified. Thus homeomorphic to the $D$-sphere $S^D$. If we assume
the ordinary round metric of the $D$-sphere, the metric under the
orthogonal projection will be 
\beq
ds^2=\frac{dr^2}{1-r^2}+r^2d\Omega_{D-1}^2
\eeq
where $d\Omega_{D-1}^2$ is the ordinary round metric of the $(D-1)$
sphere. 
The advantage of this orthogonal projection of the single scalar
field case is that if we do the projection in the scalar field
direction the hyperspheres that represent Kasner indices with value
zero become hyperplanes after the projection. The result is the
``Euclidean'' simplex inscribed in the sphere. The vacuum case can
also be ``dimensionally reduced'' in a similar way. But now it is
helpful to use the stereographic projection with respect to one of the
vertices on the sphere. The stereographic projection is illustrated in
figure \ref{stereo}. The ``null'' circles though this vertex will
again become hyperplanes, the remaining ``null'' circle will be mapped
onto a sphere, and the remaining vertices will lie on this
sphere. Again, it appears as if we have a simplex inscribed in a
sphere. However, the Euclidean metric has to replaced by the metric
\beq
ds^2=\frac{4}{(1+r^2)^2}(dr^2+r^2d\Omega_{D-2}^2)
\eeq
The last vertex of the original simplex is at infinity.
 
These ``null'' spheres will divide the space into different open
regions. Let us assume that we consider only \emph{one} of the
hemispheres in the scalar field case, their calculated volumes must
then afterwards be multiplied by 2. To determine the type of region we are considering here are
some rules (considering only vacuum case or one scalar field):
\begin{enumerate}
\item{} Each region will have a number of vertices $V>1$ connected to it.
\item{} If $E$ is the number of expanding dimensions and $C$ the number
of contracting dimensions then $V=E$ and $d=D+1=E+C$. 
\item{} The number of $V$-regions for $V\neq d$ are
$\frac{d!}{(d-V)!V!}$. 
\item{} The number of $V=d$ regions is 1 in the scalar field case and
0 in the vacuum case. 
\end{enumerate}

To calculate the various probabilities
assuming each point is equally probable we have to calculate:
\beq 
P(E+C)= \frac{\text{Vol}(E-regions)}{\text{Vol}(S^{D-1+\epsilon})}
\eeq
where $\epsilon=1$ in the presence of a scalar field and 0
otherwise. 

\begin{figure}
\centering
{\epsfig{figure=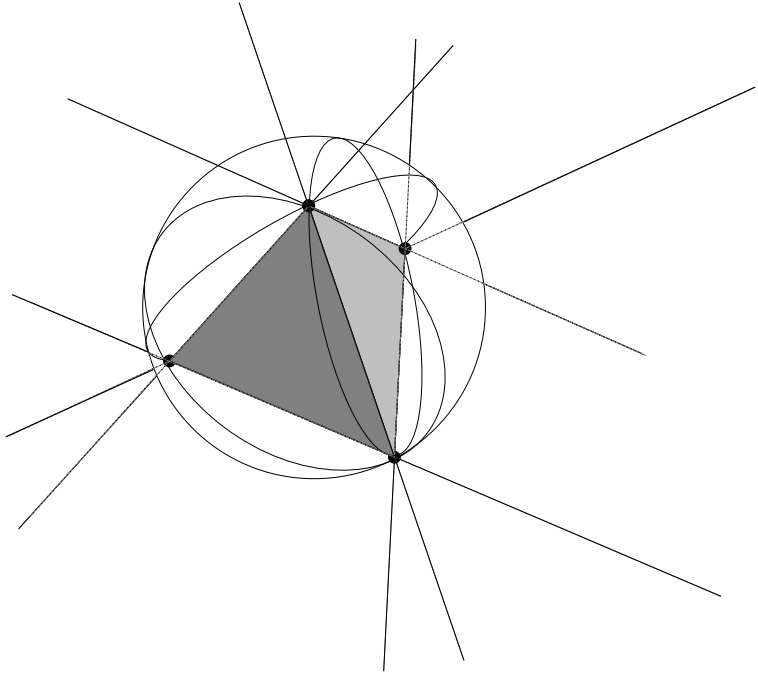, width=10cm} 
\caption{The vacuum $d=5$ case after a stereographic projection. Note that there is also a vertex at infinity.} 
\label{fivecase}}
\end{figure}

Let us look at the vacuum $d=5$ case illustrated in figure
\ref{fivecase}. The volume of the 3-sphere is Vol$(S^3)=2\pi^2$. The
region inside the tetrahedron in the center has 4 vertices, thus
corresponding to $(4e+1c)$ solutions. This tetrahedron can be
parametrized and its volume can be computed using the round
metric. Doing so, we obtain the volume:
\beq
\text{Vol}(Tetrahedron)=1.40
\eeq
(with the aid of a computer). There are in all $\frac{5!}{1!4!}=5$
such regions, thus we have:
\beq
P(4e+1c)=\frac{5\text{Vol}(Tetrahedron)}{2\pi^2}=0.355
\eeq

We can also calculate an upper bound
for $P((d-1)e+1c)$ in the vacuum case for any dimension. We note that in the vacuum case
 the central $(d-2)$-simplex in the stereographic projection has a
smaller volume than its Euclidean counterpart of the same
``size''. This calculation yields the formula:
\beq
P((d-1)e+1c)<\frac{2^{d-3}d^{\frac{d}{2}}(d-1)^{\frac{d-1}{2}}\Gamma\left(\frac{d-1}{2}\right)}{(d-2)![(d-1)^2-1]^{\frac{d-2}{2}}\pi^{\frac{d-1}{2}}}
\eeq
For $d<6$ this is actually larger than one, and is not very
informative. But in the large $d$ limit the upper bound will go as $\propto d^{-\frac{d}{2}}$
which is much smaller than if we naively would have assumed that all
the regions had equal size. In the latter case we would have got
$\approx d\cdot 2^{-d}$.


\begin{thebibliography}{99}

\bibitem{RS}
M. Ryan and L. Shepley, {\it Homogeneous Relativistic Cosmologies.}
Princeton University Press, (1975)
\bibitem{Carlip}
S. Carlip, {\it Quantum Gravity in 2+1 Dimensions}, Cambridge
University Press, (1998)
\bibitem{MyPaper}
S. Hervik, {\it Class. Quantum Grav.} {\bf 17}, 2765 (2000)
\bibitem{Nakahara}
M. Nakahara, {\it Geometry, Topology and Physics}, IoP Publishing (1990)
\bibitem{thur97}
W.P. Thurston, {\it Three-Dimensional Geometry and Topology}, 
{\bf Volume 1},  Princeton University Press (1997)
\bibitem{Polchinski}
J. Polchinski, {\it String Theory}, Cambridge University Press, (1998)
\bibitem{Green}
B. Green, {\it Lecture notes from TASI-96}, hep-th/9702155
\bibitem{LS}
V.N. Lukash and A.A. Starobinskii, {\it Sov. Phys. JETP} {\bf 39}, 742
(1974)
\bibitem{DH2000}
T. Damour and M. Henneaux, hep-th/0006171 (2000)
\bibitem{CGS}
E.J. Copeland, J. Gray and P.M. Saffin, hep-th/0003244 (2000)
\bibitem{GC}
J. Gray and E.J. Copeland, hep-th/0102090 (2001)

\end{thebibliography}
\end{document}